\documentclass[dvips,ssy,preprint]{imsart}

\RequirePackage[OT1]{fontenc}
\RequirePackage{amsthm,amsmath}
\RequirePackage[numbers]{natbib}
\RequirePackage[colorlinks,citecolor=blue,urlcolor=blue]{hyperref}

\usepackage{amsmath,amsfonts,amsthm}
\usepackage{dsfont, graphics, graphicx, color, subfigure, epsfig}

\doi{10.1214/11-SSY048}
\pubyear{2013}
\volume{3}
\issue{0}
\firstpage{1}
\lastpage{27}
\startlocaldefs
\renewcommand{\mathbb}[1]{\mathds{#1}}
\newcommand{\mathbbm}[1]{\mathds{#1}}
\newtheorem{lemma}{{Lemma}}
\newtheorem{theorem}{{Theorem}}
\newtheorem{corollary}{{Corollary}}
\newcommand{\PP}[1]{\mathbb{P}\left\{{#1}\right\}}
\newcommand{\EE}[1]{\mathbb{E}\left[{#1}\right]}
\newcommand{\II}[1]{\mathbbm 1_{\{#1\}}}
\newcommand{\ina}[1]{\mathbbm 1_{\{#1 < 0\}}|#1|}

\def\Reals{\mathbb{R}}

\def\Nats{\mathbb{N}}
\newcommand{\eref}[1]{Equation~(\ref{#1})}
\def\be{\begin{equation}}
\def\ee{\end{equation}}
\def\ben{\[}
\def\een{\]}
\def\bearn{\begin{eqnarray*}}
\def\eearn{\end{eqnarray*}}
\def\bear{\begin{eqnarray}}
\def\eear{\end{eqnarray}}
\def\barr{\begin{array}}
\def\earr{\end{array}}
\def\eps{\epsilon}

\newcommand{\lp}{\left(}

\newcommand{\rp}{\right)}

\endlocaldefs

\begin{document}

\begin{frontmatter}
\title{Stability of a stochastic model for demand-response\thanksref{Thx}}
\runtitle{Stability of a stochastic model for demand-response}
\thankstext{Thx}{This is an extended version of the conference
paper ``Satisfiability of Elastic Demand in the Smart Grid'' by the
same authors, in proceedings of the IARIA Energy 2011 conference held
in May 22-27, 2011 in Venice, Italy.}

\begin{aug}
\author{\fnms{Jean-Yves} \snm{Le Boudec}\ead[label=e1]{jean-yves.leboudec@epfl.ch}}
\and
\author{\fnms{Dan-Cristian} \snm{Tomozei}\corref{}\ead[label=e2]{dan-cristian.tomozei@epfl.ch}}

\runauthor{J.-Y. Le Boudec and D.-C. Tomozei}

\affiliation{\'Ecole Polytechnique F\'ed\'erale de Lausanne}

\address{EPFL, IC/LCA2\\
BC270 (BC Building)\\
Station 14\\
CH-1015 Lausanne\\
Switzerland\\
\printead{e1}\\
\phantom{E-mail:\ }\printead*{e2}}
\end{aug}

\begin{abstract}
We study the stability of a Markovian model of electricity production
and consumption that incorporates production volatility due to
renewables and uncertainty about actual demand versus planned
production. We assume that the energy producer targets a fixed energy
reserve, subject to ramp-up and ramp-down constraints, and that
appliances are subject to demand-response signals and adjust their
consumption to the available production by delaying their demand. When
a constant fraction of the delayed demand vanishes over time, we show
that the general state Markov chain characterizing the system is
positive Harris and ergodic (i.e., delayed demand is bounded with high
probability). However, when delayed demand increases by a constant
fraction over time, we show that the Markov chain is non-positive
(i.e., there exists a non-zero probability that delayed demand becomes
unbounded). We exhibit Lyapunov functions to prove our claims. In
addition, we provide examples of heating appliances that, when
delayed, have energy requirements corresponding to the two considered
cases.
\end{abstract}



\received{\smonth{11} \syear{2011}}

\end{frontmatter}

%\input{intro}
%s1 ###
\section{Introduction}

Recent results on modeling future electricity markets~\cite{wannegkowshameysha11b} suggest that they lead to highly undesirable equilibria for consumers, producers, or both. A main reason for such an outcome might be the combination of volatility in supply and demand, the delays required for any unplanned capacity increase, and the inflexibility of demand that leads to high disutility (cost of blackouts). Further, the use of renewable energy sources, such as wind and solar, increases volatility and worsens these effects~\cite{meynegwankowsha10}.

Demand-response is advocated~\cite{kueck2009spinning} as a mechanism to reduce ramp-up requirements and to adapt to the volatility of the electricity supply, typical of renewable sources. A deployments report~\cite{eto2007demand} shows the feasibility of delaying the starting of air conditioners by using signals from the distributor. Adaptive appliances combined with simple, distributed demand-response algorithms are advocated~\cite{keshav2010internet,maxemchuk2010}; they are assumed to reduce, or delay, their demand when the grid is not able to satisfy them. Some examples follow: e-cars, which may have some flexibility regarding the time and the rate at which their batteries can be loaded; heating systems or air conditioners, which can delay their demand if instructed to; and hybrid appliances, which use alternative sources in replacement for the energy that the grid cannot supply. If the alternative energy source is a battery, then it will need to be replenished at a later point in time, which will eventually lead to later demand.
%, which replaces the original one. %As batteries can never be 100\% efficient, the later demand is likely to be larger than the demand that it replaced.
%\end{itemize}

The presence of adaptive appliances that respond to demand-response signals helps address the volatility of renewable energy supply, however, backlogged demand is likely to be merely delayed, rather than canceled; this introduces a feedback loop into the global system of consumers and producers. Potentially, the backlogged demand can be increased to a point where future demand becomes excessive. In other words, one key question is whether it is possible to stabilize the system. This is the question we address in this paper.

To address this fundamental question, we consider a macroscopic model, inspired by the model of Meyn et al.~\cite{meynegwankowsha10}.
We assume that the electricity supply follows a two-step allocation process: First, in a forecast step (\emph{day-ahead market}) the demand and renewable supply are forecast, and the total supply is planned; Second, (\emph{real-time market}) the actual, volatile demand and renewable supply are matched as closely as possible. Like Cho and Meyn~\cite{chomey06b}, we assume that the rate at which the supply can be increased in the real-time step is subject to ramp-up and ramp-down constraints. Indeed, it is shown in this reference that it is an essential feature of the real-time market.
%This model captures the volatility of the demand and of some part of the supply, as well as the the ramp-up constraint.
We modify the model of Meyn et al.~\cite{meynegwankowsha10} and assume that the whole demand is adaptive. Although this is clearly an exaggerated assumption, we do it for simplicity and as a first step, leaving the combination of adaptive and non-adaptive demand for future research.  We are interested in simple, distributed algorithms, as suggested by Keshav and Rosenberg~\cite{keshav2010internet}, therefore, we assume that the suppliers cannot directly observe the backlogged demand; in contrast, they see only the actual instantaneous demand; at any point in time where the supply cannot match the actual demand, we assume that demand-response signals are sent and that the backlogged demand increases.

Our model is macroscopic, so we do not model in detail the mechanism by which appliances adapt to the available capacity; several possible directions for achieving this are described by Keshav and Rosenberg~\cite{keshav2010internet}. We do consider, however, two essential parameters of the adaptation process. First, the \emph{delay} $1/\lambda$ is the average delay after which frustrated demand is expressed again. Second, the \emph{evaporation} $\mu$ is the fraction of backlogged demand that disappears and will not be resubmitted per time unit. The inverse delay $\lambda$ is clearly positive; in contrast, as discussed in Section~\ref{sec-system} and in Appendix~\ref{sec-evapo}, it is reasonable to assume that some adaptive appliances naturally lead to a positive evaporation (this is the case for a simple model of heating systems), but it is not excluded that inefficiencies in some appliances lead to negative evaporation.

Within these modeling assumptions, the electricity suppliers are confronted with a scheduling problem: how much capacity should be bought in the real time market to match the adaptive demand? The effect of demand-response is to increase the latent demand, due to backlogged demand returning into the system.
%We assume that the suppliers cannot directly observe the backlogged demand; in contrast, they see only the effective instantaneous demand; at any point in time where the supply cannot match the effective demand, the backlogged demand increases.
This is the mechanism by which the system might become unstable.
We consider a threshold based mechanism~\cite{meynegwankowsha10}. It consists in targeting some fixed supply reserve at any point in time; the target reserve might not be met, due to volatility of renewable supply and of demand, and due to the ramp-up and ramp-down constraints.

Our contribution is to show that if evaporation is positive, then indeed any such threshold policy stabilizes the system. In contrast, if evaporation is negative, then there exists no threshold policy that stabilizes the system. We conjecture that when evaporation is zero, the system remains stable. %The case where evaporation is exactly equal to 0 remains unsolved.

Our results suggest that evaporation plays a central role. Simple adaptation mechanisms, as described in this paper, might work if evaporation is positive (as is generally expected), but will not work if evaporation is negative, i.e., the fact that demand is backlogged implies that a higher fraction of demand returns into the system. This suggests that future research needs to be done in order to gain a deeper understanding of evaporation, whether it can truly be assumed to be positive, and if not, how to control it.
%... other statements about sizing the backlog, effect of $\mu$, or threshold...

We use discrete time, for tractability. We use the theory of Markov chains on general state spaces~\cite{meyn-tweedie}. In Section~\ref{sec-system}, we describe the assumptions and the model, and we relate our model to prior work. In Section~\ref{sec-threshold}, we study the stability of the system under threshold policies. We give proofs in Section~\ref{sec-proofs}. In the appendix we discuss whether evaporation can be positive or negative when the appliance is a heating system, including heat pumps.

%\input{system}
%s2 ###
\section{Model and assumptions}
\label{sec-system}
%s2.1 ###
\subsection{Assumptions and notation}
We use a discrete model, where $t \in \Nats$ represents the time elapsed since the beginning of the day. The time unit represents the time scale at which scheduling decisions are made, and is of the order of the second.

The supply is made of two parts: the planned supply $G^f(t)$ (forecast in the day-ahead market), and the actual supply $G^a(t)$, which may differ, due to two causes. First, the forecasted supply might not be met, due to fluctuations, for example in wind and sunshine. Second, the suppliers attempt to match the demand by adding (or subtracting) some supply, bought in the real-time market. %As in \cite{meynegwankowsha10},
We assume that this latter term is limited by the ramp-up and ramp-down constraints. We model the actual supply as
  \begin{eqnarray}
G^a(t) & = & G(t-1) + G^f(t) + M(t),
\end{eqnarray} where $M(t)$ is the random deviation from the planned supply due to renewables, $G(t-1)$ is the supply decision in the real-time market. We view $G^{f}(t)$ as deterministic and given(it was computed yesterday in the day-ahead market), $M(t)$ as an exogenous stochastic process, imposed by nature, and $G(t-1)$ as  a control variable.

We call $D^{a}(t)$ the ``natural'' demand. It is the total electricity demand that would exist if the supply were sufficient. In addition, at every time $t$, there is the \emph{backlogged demand} $B(t)$, which results from the use of demand-response: $B(t)$ is the demand expressed at time~$t$ due to a previous demand being backlogged. The total effective demand, or expressed demand, is
 \be
  E^{a}(t)=D^a(t)+B(t).
  \ee
We model the effect of demand-response as follows:
  \begin{eqnarray}
  B(t)& = & \lambda Z(t) \label{eq-b}
 \\
 %Z(t+1) & =  & (1- \lambda -\mu) Z(t) + F(t)
 Z(t+1) & =  & Z(t)- B(t) -\mu Z(t) + F(t)
 \label{eq-z}
 \\
 F(t) & = & [E^a(t) - G^a(t)]^+. \label{eq-f}
\end{eqnarray}

In the above equations, $F(t)$ is the \emph{frustrated}
demand, i.e., that is denied satisfaction at time $t$. \eref{eq-f} expresses that, through adaptation, the demand that is served is equal to the minimum of the actual demand and the supply. The variable $Z(t)$ is the \emph{latent backlogged} demand; it is the demand that was delayed, but might be expressed later. It is incremented by the frustrated demand.

%The frustrated demand is expressed with some delay; we model this with \eref{eq-b}, where $\lambda^{-1}$ is the average delay, in time slots.
When $\mu = 0$ \eref{eq-b} can be interpreted as the approximation of a delay in the expression of frustrated demand via the use of a first-order filter (the average delay amounts to $\lambda^{-1}$ time slots).

The evolution of the latent backlogged demand is expressed by \eref{eq-z}. The returning demand $B(t)$ is removed from the backlog (some of it returns to the backlog, by means of \eref{eq-f} at some later time). The remaining backlog evaporates at a rate $\mu$, which captures the effect that delaying demand has on the total amount of demand. Delaying a demand could indeed result in a decreased backlogged demand, in which case the evaporation factor is positive. For example this occurs if we delay heating a building equipped with a heating system whose energy efficiency is constant (as we discuss in Appendix~\ref{sec-evapo-1}); such a heating system will request more energy in the future, but the integral of the energy consumed over time is less whenever some heating requests are delayed. In this case, positive evaporation comes at the expense of a (hopefully slight) decrease in comfort (measured by temperature in the room). In other cases, though, we do not exclude the possibility that evaporation be negative. This might occur for example with heat pumps, as discussed in Appendix~\ref{sec-evapo-2}.

%For example, consider an adaptive appliance that runs on a combination of grid power and battery; adaptation of the demand is made by switching to battery operation. The frustrated demand is later expressed to replenish the battery; here it might be that the integral of the energy used by such a system is larger when the battery is used, because loading and discharging a battery aseand adapt by anssum of the
%
 %$\lambda^{-1}$ is the average delay before some latent demand is expressed again.
%
%We also assumed that some latent demand evaporates; this is expressed by the factor $\mu$ in \eref{eq-z}, which is the fraction of latent demand that disappears in one time slot. The same expression can be used to capture an increase in latent demand as a consequence of backlogging, if $\mu$ is given a negative value.

Like Meyn et al.~\cite{meynegwankowsha10}, we assume that the natural demand can be forecast with some error, so that
\begin{eqnarray}
D^a(t) & = & D(t) + D^f(t),
\end{eqnarray}
where the forecasted demand $D^f(t)$ is deterministic and $D(t)$, the deviation from the forecast, is modeled as an exogenous stochastic process. We assume that the day-ahead forecast is done with some fixed safety margin $r_0$, so that
  \be
  G^f(t) = D^f(t) + r_0.
  \ee

%s2.2 ###
\subsection{The stochastic model}
We model the fluctuations in demand $D(t)$ and renewable supply $M(t)$ as stochastic processes such that their difference $M(t)-D(t)$ is the integral of an iid noise sequence $(N(t),t\ge 0)$ of zero mean $\EE N = 0$, finite second moment $\EE {N^2} = \sigma^2$, with a distribution of continuous density $\rho(\cdot)$ everywhere positive on $\mathds R$: %an ARIMA$(0,1,0)$ Gaussian process, i.e.
  \be \lp M(t+1)-D(t+1)\rp - \lp M(t) - D(t) \rp = N(t+1).
  \ee % where $N(t)$ is white Gaussian noise, with variance $\sigma^2$. This is the discrete time equivalent of Brownian motion~\cite{meynegwankowsha10}.

Let $R(t)$ be the reserve, i.e the difference between the actual production and the expressed demand, defined by
 \be
 R(t) = G^a(t) - E^a(t) %+ r_0
 \ee
and let $I(t)$ be the increment in supply bought on the real-time market, i.e.
 \be
 I(t) = G(t)-G(t-1).
 \ee

Putting all the above equations together, we obtain the system equations:
\begin{eqnarray}
R(t+1) & = & R(t) + I(t) + N(t+1) - \lambda\lp Z(t+1) - Z(t) \rp, \\
Z(t+1) & = & (1- \lambda- \mu) Z(t) + \ina {R(t)}.
\end{eqnarray}
%with $\gamma=1-\lambda-\mu$. Note that, by our system assumptions, we must have $0< \gamma <1$.
Thus we can describe our system by a two-dimensional stochastic process $X(t)=(R(t), Z(t))$, with $t \in \Nats$.

The sequence $I(t)$ is the control sequence. It satisfies the \emph{ramp-up and ramp-down constraints}:
 \ben
 -\xi \leq I(t) \leq \zeta,
 \een where $\zeta$ and $\xi$ are some positive constants.

 We assume a simple, threshold-based control that attempts to make the reserve equal to some threshold value $r^*>\zeta$; therefore
  \be
  I(t) = \max \lp \min\lp \zeta, r^* -R(t)\rp, -\xi \rp.
  \ee

%f1 ###
\begin{figure}%[ht]
%\begin{center}
\scalebox{.7}{\input{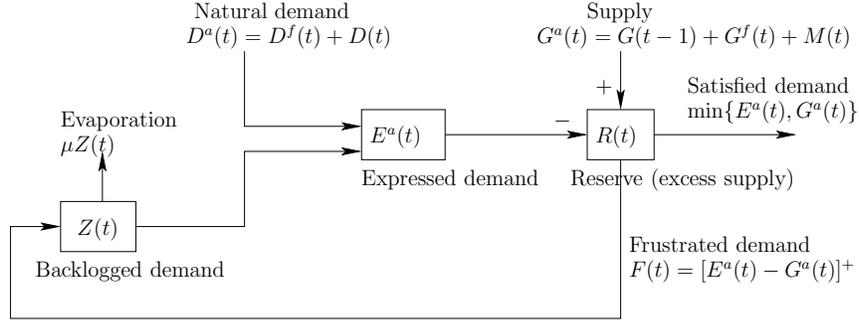}}
\caption{System model.}
\label{fig:sys}
%\end{center}
\end{figure}

In summary, we have as a model the stochastic sequence $X=(X(t))_{t\in \Nats}$ defined by
\begin{align}
\nonumber R(t+1)  = {}& R(t) - \lambda \ina{R(t)} + \lambda(\lambda+\mu)Z(t)
\\ \label{eq:el1}
&{}  + ( \zeta \wedge (r^*-R(t)) ) \vee (-\xi) + N(t+1),
\\
\label{eq:el2}
Z(t+1)  ={}  & \ina{R(t)} + (1-\lambda-\mu) Z(t),
\end{align}
where $N$ is a zero-mean finite variance iid %white Gaussian
noise sequence.  The system is depicted in Figure~\ref{fig:sys}.

Note that $X$ is a Markov chain on the state space $\textsf{S}=\Reals \times \Reals^+$.
%s2.3 ###
\subsection{Related models}
%The Gaussial Linear Model -- our model is not controllable everywhere (LSS3) and, furthermore, the eigenvalue condition (LSS5) does not hold!
Let us now discuss some similar models that have been considered in the literature.

Meyn and Tweedie~\cite{meyn-tweedie} consider the so-called Linear State Space model (LSS), which introduces an $n$-dimensional stochastic process $X = \{X_k\}_k$, with $X_k \in \Reals^n$. For matrices $F \in \Reals^{n \times n}$, $G \in \Reals^{n\times p}$, and for a sequence of i.i.d. random variables of finite mean taking values in $\Reals^p$, the process evolves as
\begin{equation}
X_k = F X_{k-1} + GW_k, \ k \ge 1.
\end{equation}
Our model~\eqref{eq:el1}-\eqref{eq:el2} can be expressed as a superposition of four such LSS models, depending on the current state of the Markov chain. The challenge of showing that our model is stable comes from the fact that in the part of the state space in which $R(t)<0$, the corresponding LSS does not satisfy the stability condition (LSS5) of~\cite{meyn-tweedie} (which requires that the eigenvalues of $F$ be contained in the open unit disk of $\mathbb C$).

Neely et al.~\cite{neely} propose a slightly different model for capturing the elasticity of demand. More specifically, the authors consider a scenario in which a deterministically {\em bounded} amount of demand arrives at each time step, and the supplier decides whether to buy an additional amount of energy from an external source at a certain cost. The unsatisfied demand is backlogged. A threshold policy is analyzed and found to be stable (the size of the backlog is found to be deterministically bounded) and optimal. Pricing decisions are also explored. The main differences with our work are the following:
\begin{itemize}
\item Additional parameters that model delay and loss of backlogged demand (i.e. $\lambda$ and $\mu$) enrich our model's expressiveness.
\item We consider potentially unbounded demand (modeled as the integral of a zero-mean finite-variance random variable), which makes proving stability more challenging.
\item No results on pricing and cost-optimality are included in the present work.
\end{itemize}

The continuous time model of Cho and Meyn~\cite{Cho05optimizationand} seeks to capture the presence of two types of energy sources, primary and ancillary, the latter being less desirable (i.e. more costly) than the former, both subject to (different) ramp-up constraints. A threshold policy is again discussed in the context of rigid demand, which is simply dropped if not enough energy is available. The analyzed Markov chain is a two-dimensional process in which the first coordinate has the quantity of energy used from the ancillary source in order to satisfy as much demand as possible, and the second coordinate has the reserve (i.e., energy surplus).

%\input{threshold}
%s3 ###
\section{System stability under a threshold policy}
\label{sec-threshold}

Let us study how the system stability of~\eqref{eq:el1}-\eqref{eq:el2} depends on the evaporation parameter $\mu$.

%Denote $\gamma := 1 - \lambda - \mu$ and
We define the following four domains:
\begin{align*}
D_1 = (-\infty, 0) \times \mathbbm R_+, \
D_2 = [0, r^* - \zeta) \times \mathbbm R_+, \
D_3 = [r^*-\zeta, r^*+\xi) \times \mathbbm R_+, \\
D_4 = [r^*+\xi, \infty) \times \mathbbm R_+.
\end{align*}
Then we can rewrite the process~\eqref{eq:el1}-\eqref{eq:el2} in matrix form:
\begin{align*}
X(t+1) = \sum_{i=1}^4\II{X(t) \in D_i} [A_iX(t) + b_i] + N_0(t+1),
\end{align*}
where
\begin{align*}
&
X(t) = \left[
\begin{array}{c}
R(t) \\
Z(t)
\end{array}
\right], \,
N_0(t) = \left[
\begin{array}{c}
N(t) \\
0
\end{array}
\right],\\
&
b_1 = b_2 = \zeta_0 = \left[
\begin{array}{c}
\zeta \\
0
\end{array}
\right], \,
b_3 = r^*_0 = \left[
\begin{array}{c}
r^* \\
0
\end{array}
\right], \,
b_4 = -\xi_0 = -\left[
\begin{array}{c}
\xi \\
0
\end{array}
\right], \,
\mbox{and}\\
&
A_1 = \left[
\begin{array}{cc}
1 + \lambda & \lambda ( \lambda + \mu ) \\
-1 & 1 - \lambda - \mu
\end{array}
\right], \,
A_2 = A_4 = \left[
\begin{array}{cc}
1 & \lambda ( \lambda + \mu ) \\
0 & 1 - \lambda - \mu
\end{array}
\right], \, \\
&
A_3 = \left[
\begin{array}{cc}
0 & \lambda ( \lambda + \mu ) \\
0 & 1 - \lambda - \mu
\end{array}
\right].
\end{align*}

The main reason the analysis of system stability is challenging is because both $A_1$ and $A_2=A_4$ admit $1$ as an eigenvalue.

%s3.1 ###
\subsection{Positive evaporation}
We first consider that the evaporation $\mu$ is positive, in other words a constant fraction of the delayed demand vanishes over time. For example, this is the setting of a heating appliance with a constant coefficient of performance, which we describe in Appendix~\ref{sec-evapo-1}.

Then we have the first result in the form of the following
\begin{theorem}
\label{thm:stab}
For some function $f \ge 1$, define the $f$-norm of a signed measure $\nu$ as $\|\nu\|_f:=\sup_{g:|g|\le f}|\nu(g)|$. If the evaporation $\mu>0$, the Markov chain~(\ref{eq:el1}-\ref{eq:el2}) is positive Harris and aperiodic. There exists a function $f \ge 1$ such that for any initial distribution $\rho$, the chain converges to its unique invariant probability measure in $f$-norm, i.e.
$$
\left\|\int \rho(dx) \mathbbm P^n(x,\cdot) - \pi(\cdot) \right\|_f \to_{n\to \infty} 0.
$$
\end{theorem}

The proof uses the theory of general state-space Markov chains. The following Lemmas are instrumental in proving the result. The proofs are found in Section~\ref{sec-proofs}.

\begin{lemma}
\label{lemma:irred}
If $1-\lambda-\mu < 1$, then there exists a measure $\varphi$ such that the Markov chain~(\ref{eq:el1}-\ref{eq:el2}) is $\varphi$-irreducible.
\end{lemma}

A set $C$ is said to be $\nu_T$-small for some non-trivial measure $\nu_T$ and a positive integer T, if for all $x\in C$, the probability of reaching any measurable $B$ is $\mathbbm P^T(x,B) \ge \nu_T(B)$.

Furthermore, a set $C$ is petite if there exists a distribution $h$ on the positive integers and a non-trivial measure $\nu_h$, such that for any $x\in C$, and for any Borel set $B$, the transition kernel of the {\em sampled chain} has the following property:
$$
K_h(x, B) := \sum_{t\ge 0} h(t) \mathbbm P^t(x, B) > \nu_h(B).
$$
A $\nu_T$-small set is implicitly $\delta_T$-petite, where $\delta_T$ is the Dirac distribution.

\begin{lemma}
\label{lemma:small}
For any set $C = J \times [0, b]$, with $J$ a finite closed interval and $b> 0$, there exists $T_0>0$ and non-trivial measures $(\nu_T)_{T\ge T_0}$ such that $C$ is $\nu_T$-small for all $T\ge T_0$.
\end{lemma}

Two direct consequences follow:

\begin{corollary}
\label{cor:aper}
The Markov chain~(\ref{eq:el1}-\ref{eq:el2}) is aperiodic.
\end{corollary}

\begin{corollary}
\label{cor:comppet}
Any compact subset of the state space is petite.
\end{corollary}

In our proof of Theorem~\ref{thm:stab}, we exhibit a Lyapunov function for the system, which has a negative drift outside of a compact set surrounding the origin. The existence of such a function, along with the previous results, enables us to apply Theorem 14.0.1 of Meyn and Tweedie~\cite{meyn-tweedie}, by which we conclude.

%s3.2 ###
\subsection{Negative evaporation}
Consider now that the evaporation $\mu$ is negative. In this case, a returning delayed job requires a higher fraction $(1-\mu)$ of resources than the original submitted job. An example of such appliance is given in Appendix~\ref{sec-evapo-2}. Then we have the following:
\begin{theorem}
\label{thm:unstab}
If $\mu<0$, the Markov chain~(\ref{eq:el1}-\ref{eq:el2}) is non-positive.
\end{theorem}

In the proof, we find a function that has finite increments and is such that, outside a certain level set, its drift is negative. By Theorem~11.5.2 from Meyn and Tweedie~\cite{meyn-tweedie} we reach the conclusion.

We sum up the two results in the following:

\begin{corollary}
If a positive fraction of the latent demand disappears during each time slot ($\mu>0$), then any threshold policy stabilizes the system. But, if delaying any job results in an increase of its requirement/workload by a positive fraction ($\mu<0$), then there exists no threshold policy that stabilizes the system.
\end{corollary}

% \input{ode}
%% \begin{lemma}
%% \label{lemma:presmall}
%% Any set $C = J \times [0, b]$, with $J$ a finite closed interval and $b> 0$, is such that there exists a measure $\varphi$ which satisfies the following property: for any Lebesgue-measurable set $B$, there exists a $T>0$ and a constant $K$ such that the probability of hitting $B$ starting from any $x\in C$ in $T$ steps is lower-bounded by $K\varphi(B)$.
%% \end{lemma}
%% \begin{proof}
%% \input{l2proof}
%% \end{proof}

%s3.3 ###
\subsection{Zero evaporation and sample trajectories}

\begin{figure}
%\begin{center}
\subfigure[$\mu = 0.2$]{\includegraphics[width=.31\textwidth]{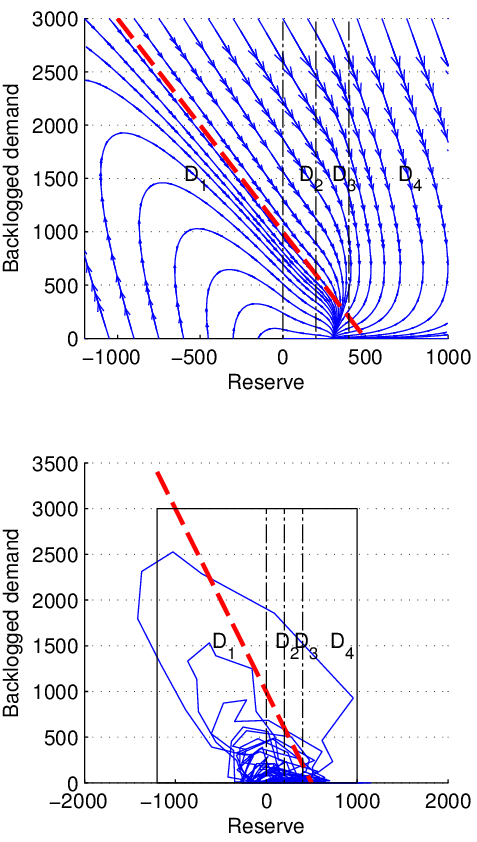} \label{fig:traj-pos}}
\subfigure[$\mu = 0$]{\includegraphics[width=.31\textwidth]{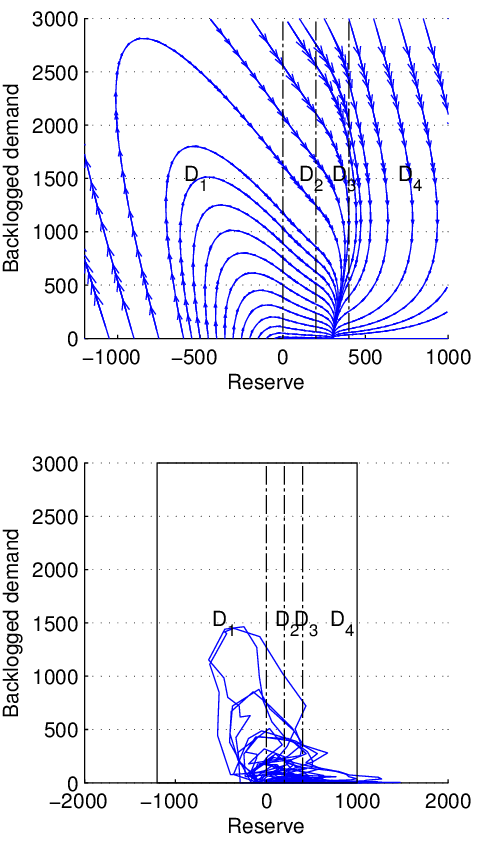} \label{fig:traj-zero}}
\subfigure[$\mu = -0.2$]{\includegraphics[width=.31\textwidth]{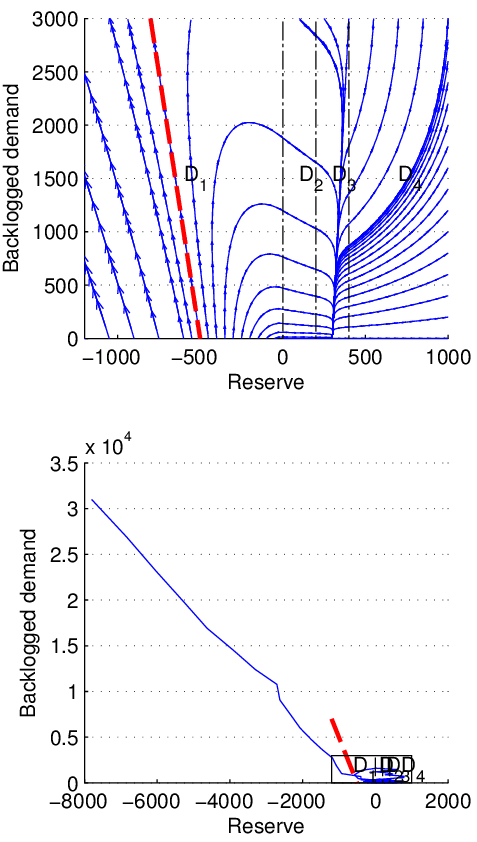} \label{fig:traj-neg}}
\caption{Fluid trajectories and $500$ iterations of the Markov process~\eqref{eq:el1}-\eqref{eq:el2} for $\zeta=\xi=100,r^*=300, \sigma=200, \lambda = 0.3$; In the top plots we plot the fluid trajectories for the rectangular regions depicted in the bottom plots.}
\label{fig:traj}
%\end{center}
\end{figure}

In Figure~\ref{fig:traj}, we plot the trajectories of the associated differential equations (or the {\em fluid trajectories}) for the three cases ($\mu>0$, $\mu = 0$, and $\mu<0$) for different starting points, along with the velocity vectors ($\dot x(t)$). These solutions can be expressed analytically and we give their expressions in Appendix~\ref{a-ode}.

%f2 ###

The limit case of $\mu = 0$ is not studied in detail this paper. We conjecture that the system remains stable in this case. The conjecture is supported by the fact that the fluid trajectories converge to the point $(r^*,0)$, just like when $\mu > 0$.

For $\mu > 0$ in region $D_1$ all the fluid trajectories have an asymptote given by the following line, shown in Figure~\ref{fig:traj-pos}:
\begin{equation}
\label{eq:asmpt}
(\lambda+\mu)z + r = \frac\zeta\mu.
\end{equation}

When $\mu < 0$ and $\lambda+\mu>0$ in region $D_1$ the fluid trajectories will diverge to minus infinity if the starting point is {\em below} the same line \eqref{eq:asmpt} (shown in Figure~\ref{fig:traj-neg}). If the starting point is above this line, the $r(t)$ coordinate will tend to $+\infty$, thus arriving in the $D_2$ domain, where the trajectory is again ``well-behaved''.

In addition to the fluid trajectories, in Figure~\ref{fig:traj} we also plot a sample trajectories of the Markov chain in the three settings. We notice that for $\mu \ge 0$ most of the time the backlogged demand is small. However excursions in the region of large backlogged demand and negative reserve occur (they are not rare events).

%s4 ###
\section{Proofs}\label{sec-proofs}

We now proceed to prove the results stated Section~\ref{sec-threshold}. Note that we consider both ramp-up and ramp-down constraints. In a previous work~\cite{iaria}, we considered only the ramp-up constraint.

Let $\gamma := 1 - \lambda - \mu$ and note that $\gamma<1$ whenever $\mu >0$.
%s4.1 ###
\subsection{Proof of Lemma~\ref{lemma:irred}}

%\input{irredproof}
%\begin{proof}
Fix some finite closed interval $I$ and some $a > 0$, and consider the set $E = I \times [0, a]$.

Consider the measure $\varphi_E$ defined as follows: for any Borel set $A$,
\begin{equation}
\label{eq:phidef}
\varphi_E(A) := \nu (A \cap E),
\end{equation}
where $\nu$ denotes the Lebesgue measure on $\mathbbm R^2$. Take a measurable set $B \subset E$. We denote the support sets of $B$ on the two dimensions by
$$
\begin{array}{l}
B_1 := \{r \in \mathbbm R: \exists \ z, \mbox{ such that } (r,z) \in B\}, \\
B_2 := \{z \in \mathbbm R_+: \exists \ r, \mbox{ such that } (r,z) \in B\}.
\end{array}
$$
We further denote the cross-section of $B$ at some $z \in \mathbbm R_+$ by
$$
\pi_{[z]}B := \{r \in \mathbbm R : (r,z) \in B\}.
$$

Consider any initial point $x = (r,z)$. We show that whenever $\varphi_E(B)$ is strictly positive, there exists a finite $T$ such that the probability of hitting $B$ in $T$ steps starting from $x$ is also strictly positive. Then, by Proposition 4.2.1 (ii) from Meyn and Tweedie~\cite{meyn-tweedie}, we conclude.

Let us thus consider $\varphi_E(B) > 0$. For any $0<\epsilon<1$, there exists a $\delta>0$, such that the Lebesgue measure of the set $B^\delta := \{(r,z)\in B \cap E : z \ge \delta\} = B \cap E \cap (\mathbbm R \times [\delta, \infty))$ is strictly positive and, furthermore,
\begin{equation}
\label{eq:epsfrac}
\epsilon = \frac {\varphi_E(B^\delta)} {\varphi_E(B)} = \frac {\nu(B^\delta)} {\varphi_E(B)}.
\end{equation}
%can be expressed as a fraction $\epsilon$ of the measure of $B$: $\varphi_E(B^\delta) = \epsilon \varphi_E(B) > 0$. Indeed, the function $\delta \mapsto f(\delta) := \varphi_E( B^\delta )$ is continuous. Hence, it takes all the values between $f(0) = \varphi_E(B)$ and $f(a) = 0$. Furthermore, since $B^\delta \subset E$, we have that $\varphi_E(B^\delta) = \nu(B^\delta)$.

For every $0 < \alpha \le 1$, we define the sets $B^\delta(\alpha) := \{(r,z) : (r, \alpha z) \in B^\delta\}$. Clearly,
\begin{equation}
\label{eq:scale}
\nu(B^\delta(\alpha)) = \frac {\nu(B^\delta)} \alpha = \frac {\epsilon\varphi_E(B)} \alpha,
\end{equation}
and, furthermore,
\begin{equation}
\label{eq:proj}
\pi_{[z]}B^\delta(\alpha\beta) = \pi_{[\beta z]}B^\delta(\alpha), \ \forall \ 0 < \alpha, \beta \le 1.
\end{equation}

Assume that the initial point $x = (r,z)$ is in $D_1$.

%f3 ###
\begin{figure}[t]
%\begin{center}
\resizebox{4in}{!}{\input{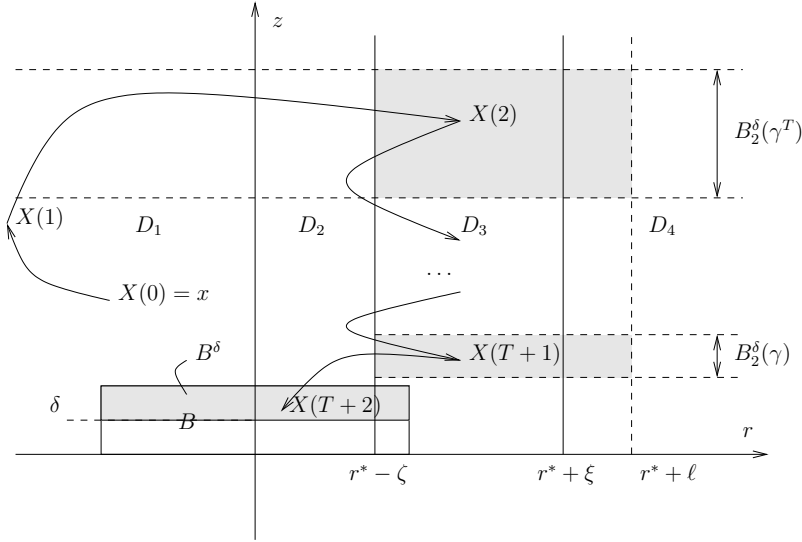}}
\caption{Typical trajectory.}
\label{fig:proof}
%\end{center}
\vspace*{-9pt}
\end{figure}

We assess the probability of the particular trajectory from $x$ to $B$, depicted in Figure~\ref{fig:proof} (more specifically to $B^\delta$). This probability lower bounds the probability of reaching $B$. Specifically, we determine a sufficient value of $T$ such that
\begin{enumerate}
\item $X(1) \in D_1$,
\item $X(2),\dots, X(T+1) \in D_3\cup D_4$, and
\item $X(T+2) \in B^\delta$.
\end{enumerate}
\eject

Let thus $(R(0),Z(0)) = (r,z)$ and denote
\begin{align*}
r(1) &= (1+\lambda)r + \lambda(1-\gamma)z + \zeta,\\
z(1) &= -r+\gamma z.
\end{align*}
We have that $R(1) = r(1) + N(1)$ and $Z(2) = \gamma z(1) - r(1) - N(1)$.

We select a large enough $T$ such that the event $\omega_1 = \{N(1) < -r(1)\}$ includes the event $\omega_2 = \{Z(2) \in B_2^\delta(\gamma^T)\}$. We have that $\omega_2 \subset \{ \gamma z(1) - \sup B_2^\delta(\gamma^T) < r(1) + N(1) < \gamma z(1) - \inf B_2^\delta(\gamma^T)\}$. Since $\inf B_2^\delta(\gamma^T) \ge \frac \delta {\gamma^T}$, it is sufficient to consider some $T$ such that
$\frac \delta {\gamma^T} \ge \gamma z(1)$. We choose a $T$ such that $T \ge T_0(B, \epsilon, x) \wedge 0$, where
\begin{equation}
\label{eq:tbound}
T_0(B, \epsilon, x) = \left \lceil \frac 1 {\log \gamma} \log \frac \delta {z(1)} \right \rceil.
\end{equation}
For such $T$, if we consider $N(1)$ such that $\omega_2$ occurs, then necessarily $\omega_1$ occurs, and $R(1) < 0$. In other words, at the beginning of time step $2$ the process is still in region $D_1$.

For $t=2, \dots, T+1$, we consider $N(t)$ such that $R(t) \in L(\ell)$, where
$$
L(\ell) = (r^* - \zeta, r^* + \ell]
$$
for a fixed $\ell > \xi \wedge \zeta$. As for such $t$ the process evolves in domains $D_3$ and $D_4$, the evolution of $Z(t)$ will be purely deterministic and can be written explicitly after $2<t\le T+2$ steps as $Z(t) = \gamma^{t-2} Z(2)$. By definition, since $Z(2)\in B_2^\delta(\gamma^T)$, we have that $Z(T+2) \in B_2^\delta$. We subsequently consider $N(T+2)$ such that $(R(T+2), Z(T+2))\in B^\delta$, or, equivalently, $(R(T+2), Z(2)) \in B^\delta(\gamma^T)$.

Let us now compute a lower bound of the probability of this trajectory. For some set $A\subset \mathbbm R$ and some real number $b \in \mathbbm R$, we denote $-A := \{-a : a \in A \}$ and $A+b := \{a+b : a \in A \}$.

We can now characterize the probability of hitting $B$ in $T+2$ steps following the trajectory that we just described:
\begin{align*}
\mathbbm P^{T+2}(x,B) &\ge \int_{(\gamma z(1) - B_2^\delta(\gamma^T)) \times \{z(1)\}} {\mathbbm P( x, dx_1 )} \int_{L(\ell) \times B_2^\delta(\gamma^T)} {\mathbbm P( x_1, dx_2 )} \\
& {\int_{L(\ell) {\times} B_2^\delta(\gamma^{T-1})}} {\mathbbm P( x_2, dx_3 )}
{\dots} {\int_{L(\ell) {\times} B_2^\delta(\gamma)}} {\mathbbm P( x_T, dx_{T{+}1} )} \mathbbm P( x_{T{+}1}, B^\delta ).
\end{align*}
Recall that the density of $N$, $\rho(\cdot)$ is continuous and strictly positive everywhere.

The probability that $N(1)$ is such that $Z(2) \in B_2^\delta(\gamma^T)$ (i.e., event $\omega_2$) can be written as
$$
\PP{\omega_2} = \int_{\gamma z(1) - r(1) - B_2^\delta(\gamma^T)} \rho(dn_1).
$$

Subsequently, for all $2\le t\le T+1$, the probability that $N(t)$ is such that $R(t) \in L(\ell)$ is always strictly positive (because the transition occurs between points belonging to bounded sets, and the density $\rho$ is strictly positive everywhere). A positive lower bound on this probability depends on the starting point $x$, the target set $B$, the number of steps $T$, and the considered $\epsilon$. We denote it by $p(T, B, \epsilon, x)$.

In the last step we want to ensure that $R(T+2)$ belongs to the corresponding cross-section $\pi_{[\gamma^T(\gamma z(1) - r(1) - n_1)]}B^\delta$, which by~(\ref{eq:proj}) is identical to $\pi_{[\gamma z(1) - r(1) - n_1]}B^\delta(\gamma^{T})$. By the same argument as above, there exists $\eta = \eta(T,B,\epsilon,x) > 0$ such that
$$
\int_{\gamma z(1) - r(1) - B_2^\delta(\gamma^T)} \rho(dn_1) \int_{\pi_{[\gamma z(1) - r(1) - n_1]}B^\delta(\gamma^T)}\rho(dn_{T+2}) \ge \eta \nu(B^\delta(\gamma^T)).
$$

We can then conclude by~(\ref{eq:scale}):
\begin{equation}
\label{eq:detlb}
\mathbbm P^{T+2}(x,B) \ge \frac {\epsilon \eta p^T } {\gamma^T} \varphi_E(B).
\end{equation}

When the initial point $x$ is in $D_2$, $D_3$, or $D_4$, we just need to consider $N(1)$ such that
$
(R(1), Z(1))\in D_1
$ and $Z(2) = \gamma^2 z - r(1) - N(1) \in B_2^\delta(\gamma^T)$, for some $T$ greater than a suitably chosen $T_0$, where
$$
r(1) =
\left\{
\begin{array}{ll}
r + \zeta + \lambda(1-\gamma)z, & \mbox{for $x\in D_2$},\\
r^* + \lambda(1-\gamma) z, & \mbox{for $x\in D_3$}, \\
r - \xi + \lambda(1-\gamma)z, & \mbox{for $x\in D_4$}.
\end{array}
\right.
$$
The rest of the analysis follows.
\endproof

%%% Local Variables:
%%% mode: latex
%%% TeX-master: "grid"
%%% End:

%s4.2 ###
\subsection{Proof of Lemma~\ref{lemma:small}}

%\input{smallproof}
%\begin{proof}
Assume without loss of generality that $C \cap D_1$ has positive Lebesgue measure. Let us show that the set $C_1 = C \cap D_1$ satisfies the ``smallness'' property in the statement.

In Lemma~\ref{lemma:irred}, we essentially proved that any bounded
Lebesgue-measurable set $B$ of positive measure is reachable from any
state $x$ in a finite number of steps. However, in order to give an
upper bound on the required number of steps, we defined the set $E = I
\times [0,a]$ and we introduced the measure $\varphi_E$, which is
defined as the Lebesgue measure of the set obtained via intersection
with $E$. We obtained that for any $0 < \epsilon \le 1$, for any $B$
such that $\varphi_E(B) > 0$, and for any $x$, there exists
$T_0(B,\epsilon,x) > 0$, such that for any $T \ge T_0(B, \epsilon,
x)$, there exists a strictly positive constant $K = K(T, B, x) > 0$
such that
\begin{equation}
\label{eq:lb}
\mathbbm P^{T}(x,B) \ge K( T, B, x ) \epsilon \varphi_E(B).
\end{equation}

In order to prove smallness, we need to eliminate the dependence of $K$ and $T_0$ on the specific starting point $x$ and on the destination $B$.

The dependence on $x$ can easily be dealt with. Fix some destination set $B$. Since the set $C_1$ is bounded, we can prove that
\begin{itemize}
\item $T_0(B, \epsilon, C_1 ) := \sup_{x\in C_1} T_0(B, \epsilon, x )$ is finite and
\item $K( T, B, C_1 ) := \inf_{x\in C_1} K( T, B, x )$ is strictly positive.
\end{itemize}
Indeed, revisiting the proof of Lemma~\ref{lemma:irred}, we have
\begin{equation}
\label{eq:tboundnox}
T_0(B, \epsilon, C_1) = \sup_{x\in C_1} \left \lceil \frac 1 {\log \gamma} \log \frac \delta {z(1)} \right \rceil %= \left \lceil \frac 1 {\log \gamma} \log \frac \delta {b+\gamma|\inf I|} \right \rceil
< \infty.
\end{equation}
Furthermore, $K = \eta p^T \gamma^{-T}$, and by the fact that $C_1$ is bounded we have that
\begin{align*}
p(T, B, \epsilon, C_1) &:= \inf_{x\in C_1} p(T, B, \epsilon, x) > 0, \\
\eta(T,B,\epsilon, C_1)&:= \inf_{x\in C_1} \eta(T, B, \epsilon, x) > 0.
\end{align*}

The dependence on the destination set $B$ is more tricky. The major inconvenience is the dependence on $B$ of the proposed value for $T_0$, which, as seen in~(\ref{eq:tboundnox}), depends logarithmically on $\delta$. Hence, for any $\epsilon$, we can always find a set $B$ which is such that $\delta$ needs to be arbitrarily close to $0$ in order to satisfy~(\ref{eq:epsfrac}). This in turn leads to $T_0$ being arbitrarily large (whereas we would like it to be constant). This is due to the deterministic geometric convergence of $Z(t)$ toward $0$ in region $D_3$. In other words, if starting from a strictly positive value, $Z(t)$ cannot reach $0$ in a finite time following the trajectory we considered.

Let us instead fix $\delta$ and choose a finite closed interval $I$ and constant $\Delta >\delta$. We define the set $F := I\times [\delta, \Delta]$ and consider the measure $\varphi_F$, such that $\varphi_F(A) = \nu(A\cap F)$, like in \eqref{eq:phidef} (recall that $\nu$ is the Lebesgue measure). Clearly $\nu(F^\delta) = \nu(F)$, so in this case we are allowed to choose $\epsilon = 1$ (i.e., we can evaluate the probability of reaching $B = B^\delta$), and thus the issue of $T_0$ is solved. The dependence on $B$ of $\eta$ and $p$ can be removed by considering again an infimum and by using the fact that the sets $F$ and $L(\ell)$ are bounded.

With $T_0 = T_0(F, 1, C_1)$ given by~(\ref{eq:tboundnox}), but for the chosen $\delta>0$ (no dependence on any specific destination set), for all $T\ge T_0$ we define the measure
$
\nu_T(\cdot) := \frac {\eta p^T } {\gamma^T} \varphi_F(\cdot).
$
By~(\ref{eq:detlb}) we can conclude that $C_1$ is $\nu_T$-small.

In a similar way, we can show that the entire set $C$ is small.
\endproof %\end{proof}

%%% Local Variables:
%%% mode: latex
%%% TeX-master: "grid"
%%% End:

%s4.3 ###
\subsection{Proof of Theorem~\ref{thm:stab}}
%\input{lyapunov}
%\begin{proof}
%Let us give an outline of the proof.
In the following we define a Lyapunov function $H$ which is {\em unbounded off petite sets}, that is for any $n < \infty$ the sublevel set $C_H(n) := \{y : H(y) \le n\}$ is small. We show that there exist constants $a, b, c>0$, a set $C = [-a,a] \times [0,b]$, and a function $f\ge 1$ such that the drift satisfies
$$
\mathcal D H(x) := \mathbbm E_x H( X(1) ) - H(x) \le -f(x) + c \mathbbm 1_{C}.
$$
A consequence of this result is that the chain is Harris recurrent, by Theorem 9.1.8 of Meyn and Tweedie~\cite{meyn-tweedie} and by Corollary~\ref{cor:comppet} (which shows that the set $C$ is small). Furthermore, by Theorem 10.0.1 of Meyn and Tweedie~\cite{meyn-tweedie}, it admits a unique invariant measure $\pi$. Finally, by Theorem 14.0.1 of Meyn and Tweedie~\cite{meyn-tweedie} and by Corollary~\ref{cor:aper}, we get the finiteness of $\pi$ (and thus the $f$-ergodicity) and we conclude.

Let us now delve into the details:

We prove that the function $H : \mathbbm R \times \mathbbm R_+ \to \mathbbm R_+$,
\begin{equation}
\label{eq:lyapunov}
x = \left[\begin{array}{l} r \\ z\end{array}\right] \mapsto
H(x) = %\left\{
%\begin{array}{ll}
( r + \lambda z )^2 + ( r + ( \lambda + \mu ) z )^2 %& \mbox{if $\mu > 0$, $r<0$} \\
%% -\frac 1 \zeta  (r+\lambda z) & \mbox{if $\mu = 0$, $r < -\lambda z$} \\
%% \frac 1 \zeta \left|r - \frac {\lambda^2 - \lambda + 1}{1-\lambda}z\right|& \mbox{if $\mu = 0$, $-\lambda z \le r < 0$} \\
%% ( r + \lambda z )^2 + z^2 & \mbox{if $0 \le r < r^* - \zeta$} \\
%% ( (1-\lambda-\mu) r - \lambda(\lambda+\mu) z )^2 + z^2 & \mbox{if $r \ge r^*-\zeta$}
%% \end{array}
%% \right.
\end{equation}
is a Lyapunov function for the system~(\ref{eq:el1}-\ref{eq:el2}).

For each region $D_i$ we proceed as follows. We find a suitable coordinate change $M_i$, which diagonalizes matrix $A_i = M_i \Lambda_i M_i^{-1}$ (where $\Lambda_i = \mbox{diag}(\alpha_{i1},\alpha_{i2})$ is a diagonal matrix containing the eigenvalues  of $A_i$). Specifically, we consider
$$
Y_i(t) = \left[
\begin{array}{c}
U_i(t) \\
V_i(t)
\end{array}
\right]
= M_i^{-1}X(t),
$$
along with the region constraints
$$
Y_i(t) \in \Delta_i := \{y\in \mathds R^2 : M_iy \in D_i\}.
$$
We obtain independently evolving coordinates $U_i$ and $V_i$.
This facilitates determining the set $C_i = D_i\cap C$, as well as the value of the function $f|_{D_i}$ in this region, which is such that $f|_{D_1}(x) \ge 1$ for all $x \in D_i \setminus C$.

We now proceed to determine $C_i$ and $f|_{D_i}$ for $i=1,2,3,4$.

%Now consider the case $X(t) \in D_1$ and $\mu > 0$.

Matrix $A_1$ can be diagonalized as $A_1 = M_1 \Lambda_1 M_1^{-1}$, where:
$$
M_1 =
\left[
\begin{array}{cc}
-\lambda-\mu & -\lambda\\
1 & 1
\end{array}
\right], \
M_1^{-1} =
\frac 1 \mu
\left[
\begin{array}{cc}
-1 & -\lambda\\
1 & \lambda+\mu
\end{array}
\right], \
\Lambda_1 =
\left[
\begin{array}{cc}
1 & 0 \\
0 & 1-\mu
\end{array}
\right].
$$
Consider the initial point $X(0) = x = (r,z)\in D_1$. Then, in the coordinates $(U_1,V_1)$, we have
\begin{equation}
\label{eq:chvar}
\left\{
\begin{array}{l}
U_1(1) = U_1(0) - \frac 1 \mu (N(1) + \zeta), \\
V_1(1) = (1-\mu)V_1(0) + \frac 1 \mu (N(1) + \zeta),
\end{array}
\right.
\end{equation}
under the constraints
\begin{align}
Y_1(0) = M_1^{-1}x = \mu^{-1}
\left[
\begin{array}{c}
-(r + \lambda z)\nonumber\\
r + (\lambda + \mu)z
\end{array}
\right]
\in \Delta_1, \nonumber\\
\Delta_1 = \{ (u, v) : u + v \ge 0, \ (\lambda+\mu)u + \lambda v > 0 \}.
\label{eq:delta1}
\end{align}

Hence, the Lyapunov function can be rewritten as
\begin{align*}
W_1:\Delta_1 \to \mathbbm R_+, \ W_1( y ) := H( M_1y ) = \mu^2 ( u^2 + v^2 ).
\end{align*}

\begin{figure}[t]
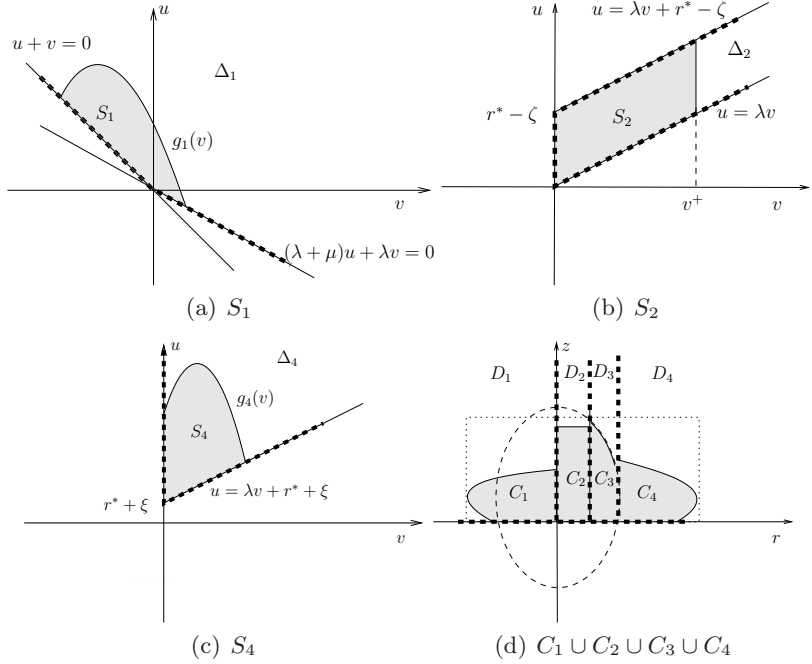

%\begin{center}
\subfigure[$S_1$]{\label{fig:insdom1}\resizebox{!}{1.5in}{\input{inst.pstex_t}}}
\subfigure[$S_2$]{\label{fig:insdom2}\resizebox{!}{1.5in}{\input{inst2.pstex_t}}}
\subfigure[$S_4$]{\label{fig:insdom4}\resizebox{!}{1.5in}{\input{inst4.pstex_t}}}
\subfigure[$C_1 \cup C_2 \cup C_3 \cup C_4$]{\label{fig:insdom3}\resizebox{!}{1.5in}{\input{inst3.pstex_t}}}
\caption{Positive drift sets.}
%\end{center}
\end{figure}

Let us compute the drift of $H$ in $x\in D_1$, $\mathcal D
H(x) \equiv \mathcal D W_1(y)$:
\begin{align*}
\mathcal D W_1(y) & = \mathbbm E_y(W_1(Y(1))) - W_1(y) \\
& = -\mu^3(2-\mu) v^2 + 2\zeta\mu(1-\mu)v -2\zeta\mu u + 2(\zeta^2 + \sigma^2).
\end{align*}
Since $\mu>0$, the set $\bar S_1$ of $y \in \Delta_1$ for which $\mathcal D W_1(y) \le -1$ is given by the hypergraph of the second degree polynomial function
$$
g_1(v) = - \frac {\mu^2(2-\mu)} {2\zeta} v^2 + (1-\mu)v + \frac {2(\zeta^2+\sigma^2) + 1} {2\zeta\mu},
$$
namely $\bar S_1 = \mbox{hyp}^+(g_1) \cap \Delta_1 = \{(u,v) \in \Delta_1 : u \ge g_1(v) \}$. The complementary set $S_1$ on which $\mathcal D W_1(y) > -1$ is $S_1 = \mbox{hyp}^- (g_1) \cap \Delta_1$. We represent both sets in Figure~\ref{fig:insdom1}. In the original coordinates, we find $C_1 = M_1 S_1 = \{M_1 y : y \in S_1\}$.

It follows that $f$ is quadratic in $r$ and $z$ in region $D_1$:
\begin{align*}
f|_{D_1}(x) ={} & \max\{1, \mu(2-\mu) (r+(\lambda+\mu)z)^2 - 2\zeta(1-\mu)(r+(\lambda+\mu)z) \\
&{} -2\zeta (r+\lambda z) - 2(\zeta^2 + \sigma^2)\}.
\end{align*}

%f4 ###

Matrix $A_2$ can be diagonalized as $A_2 = M_2 \Lambda_2 M_2^{-1}$, where
$$
M_2 =
\left[
\begin{array}{cc}
1 & -\lambda\\
0 & 1
\end{array}
\right], \
M_2^{-1} =
\left[
\begin{array}{cc}
1 & \lambda\\
0 & 1
\end{array}
\right], \
\Lambda_2 =
\left[
\begin{array}{cc}
1 & 0 \\
0 & 1-\lambda-\mu
\end{array}
\right].
$$
Consider the initial point $X(0) = x = (r,z) \in D_2$. Then, in the coordinates $(U_2,V_2)$ we have
\begin{equation}
\label{eq:chvar2}
\left\{
\begin{array}{l}
U_2(1) = U_2(0) + N(t+1) + \zeta, \\
V_2(1) = (1-\lambda-\mu)V_2(0),
\end{array}
\right.
\end{equation}
under the constraints
$$
Y_2(0) = M_2^{-1}x =
\left[
\begin{array}{c}
r+\lambda z \\
z
\end{array}
\right]
\in \Delta_2 = \{(u,v) : 0\le u - \lambda v < r^* - \zeta, \ v \ge 0\}.
$$

The Lyapunov function~(\ref{eq:lyapunov}) can be rewritten as
$$
W_2:\Delta_2 \to \mathbbm R_+, \  W_2(y) := H( M_2 y ) = u^2 + (u+\mu v)^2.
$$
Then, the drift of $H$ in $x \in D_2$ can be written as $\mathcal D H(x) \equiv \mathcal D W_2(y)$:
\begin{align*}
\mathcal D W_2(y) ={} & \mathbbm E_y(W_2(Y(1))) - W_2(y) = - \mu^2(2-(\lambda+\mu))(\lambda+\mu) v^2 \\
&{} - 2\mu(\lambda+\mu) uv + 4\zeta u + 2\mu(1-\lambda-\mu)\zeta v + 2\zeta^2 + 2\sigma^2.
\end{align*}
Since $\lambda v \le u < \lambda v + r^* - \zeta$, we have that
$$
\mathcal D W_2(y) \le 2 \sigma^2 - 2\zeta^2 + 4\zeta r^* + 2\zeta[2\lambda + \mu(1-\lambda-\mu)]v - \mu(\lambda+\mu)^2(2 - \mu)v^2.
$$
Hence, there exists a $v^+$ such that $\mathcal D W_2(y) \le -1$ for $v > v^+$. Denote by $S_2 = \{ y = (u,v) \in \Delta_2: v < v^+ \}$ (which we represent in Figure~\ref{fig:insdom2}). We obtain the corresponding set in the original coordinates: $C_2 = M_2 S_2 = \{ M_2 y : y \in S_2 \}$.

We obtain that $f$ is quadratic in $z$ in region $D_2$:
\begin{align*}
f|_{D_2}(x) = {}&\max\{1,-2 \sigma^2 + 2\zeta^2 - 4\zeta r^* - 2\zeta[2\lambda + \mu(1-\lambda-\mu)]z \\
&{}+ \mu(\lambda+\mu)^2(2 - \mu)z^2\}.
\end{align*}

Since $A_4 = A_2$, we have $M_4 = M_2$. We consider an initial point $X(0) = x = (r,z) \in D_4$. In the coordinates $(U_4,V_4)$ we have
\begin{equation}
\label{eq:chvar4}
\left\{
\begin{array}{l}
U(t+1) = U(t) + N(t+1) - \xi, \\
V(t+1) = (1-\lambda-\mu)V(t).
\end{array}
\right.
\end{equation}
under the constraints
$$
Y_4(0) = M_2^{-1} x = \left[
\begin{array}{c}
r+\lambda z \\
z
\end{array}
\right]
\in
\Delta_4 = \{(u,v) : r^* + \xi \le u - \lambda v, \ v \ge 0\}.
$$

We write the Lyapunov function~(\ref{eq:lyapunov}) in the new coordinates:
$$
W_4:\Delta_4 \to \mathbbm R_+, \  W_4(y) := H( M_2 y ) = u^2 + (u+\mu v)^2.
$$
Then, the drift of $H$ in $x \in D_4$ can be written as $\mathcal D H(x) \equiv \mathcal D W_4(y)$
\begin{align*}
\mathcal D W_4(y) ={} & \mathbbm E_y(W_4(Y(1))) - W_4(y) = - \mu^2(2-(\lambda+\mu))(\lambda+\mu) v^2 \\
&{} - 2\mu(\lambda+\mu) uv - 4\xi u - 2\mu(1-\lambda-\mu)\xi v + 2\xi^2 + 2\sigma^2.
\end{align*}

Since $u \ge \lambda v + r^* + \xi$, we have that
$$
\mathcal D W_4(y) \le 2\sigma^2 + 2\xi^2 -4\xi u - (2\mu(\lambda+\mu)r^* + 2\xi\mu)v - \mu(\lambda+\mu)^2(2-\mu)v^2.
$$
Then, we have that $\mathcal D W_4(y) < -1$ for points $y = (u,v)$ in the hypergraph of the second degree concave polynomial function
$$
g_4(v) = \frac 1 {4\xi} \left[2\sigma^2 + 2\xi^2 + 1 - (2\mu(\lambda+\mu)r^* + 2\xi\mu)v -
\mu(\lambda+\mu)^2(2-\mu)v^2\right].
$$
In Figure~\ref{fig:insdom4} we represent the set $S_4$ on which $\mathcal D W_4(y) > -1$, namely $S_4 = \mbox{hyp}^- (g_4) \cap \Delta_4$. In the original coordinates, we get $C_4 = M_2 S_4 = \{M_2 y : y \in S_4\}$.

The function $f$ is quadratic in $z$ and linear in $r$ in domain $D_4$:
\begin{align*}
f|_{D_4}(x) ={} & \max\{1, -2\sigma^2 - 2\xi^2 +4\xi (r+\lambda z) + (2\mu(\lambda+\mu)r^* + 2\xi\mu)z \\
&{} + \mu(\lambda+\mu)^2(2-\mu)z^2 \}.
\end{align*}

Finally, we write the drift of $H$ in some $x = (r,z) \in D_3$:
\begin{align*}
\mathcal D H(x) ={}& \mathbbm E(\lambda z + r^* + N)^2 + \mathbbm E((1-\mu)(\lambda + \mu)z + r^* + N)^2 \\
&{} - (r+\lambda z)^2 - (r+(\lambda+\mu)z)^2.
\end{align*}
We have that
$$
\mathcal D H(x) \le 2({r^*}^2 + \sigma^2 - r^2) + 2r^*(\lambda + (\lambda + \mu)(1-\mu))z - (\lambda+\mu)^2\mu(2-\mu)z^2.
$$
Thus, $\mathcal D H(x) \le -1$ holds for points $x = (r,z) \in D_3$ found outside the ellipse defined by
$$
\mathcal E = \{(r,z) : 2r^2 + \alpha(z - \beta \alpha^{-1})^2 = 1 + 2{r^*}^2 + 2\sigma^2 + {\beta^2} \alpha^{-1} \},
$$
where we denoted $\alpha = (\lambda+\mu)^2\mu(2-\mu)$ and $\beta = r^*(\lambda + (\lambda + \mu)(1-\mu))$.
We find $C_3 = \mbox{int} (\mathcal E) \cap D_3$.

The function $f$ is quadratic in $r$ and $z$ in domain $D_3$:
$$
f|_{D_3} = \max\{1,2(r^2 - {r^*}^2 - \sigma^2) - 2r^*(\lambda + (\lambda + \mu)(1-\mu))z + (\lambda+\mu)^2\mu(2-\mu)z^2\}.
$$

We have found that $\mathcal D H \le -f \le -1$ outside the bounded compact set $C_1 \cup C_2 \cup C_3 \cup C_4$, which concludes the proof.

%\end{proof}
\endproof

%%% Local Variables:
%%% mode: latex
%%% TeX-master: "grid"
%%% End:

%s4.4 ###
\subsection{Proof of Theorem~\ref{thm:unstab}}
%\input{unstabproof}
%\begin{proof}
Notice that if $\mu\le-\lambda$, then the $Z$ coordinate of the Markov chain cannot decrease. Hence the chain is trivially unstable (it is not even $\psi$-irreducible).

In the case $-\lambda < \mu < 0$, we turn again to Meyn and Tweedie~\cite{meyn-tweedie} to prove the theorem statement. In this case the Markov chain~(\ref{eq:el1}-\ref{eq:el2}) is $\psi$-irreducible. We need to find a function $H$ that satisfies the hypothesis of Theorem~11.5.2 from Meyn and Tweedie~\cite{meyn-tweedie} to show non-positivity.

Like in the proof of Theorem~\ref{thm:stab}, consider the change of coordinates $Y_1(t) = M_1^{-1} X(t)$, which yields evolution~(\ref{eq:chvar}) for $Y_1(t) = (U_1(t), V_1(t)) \in \Delta_1$ defined in~(\ref{eq:delta1}). The state space under these new coordinates becomes $\Delta = \{(u,v) : u+v \ge 0\}$.

Define $H : \Delta \to \mathbbm R_+$,
\begin{equation}
\label{eq:lyapunov2}
y = \left[\begin{array}{l} u \\ v\end{array}\right] \mapsto
H(y) = \left\{
\begin{array}{ll}
\log ( v ) & \mbox{if $v \ge 1$,} \\
0 & \mbox{otherwise.}
\end{array}
\right.
\end{equation}

Let us show that $H(Y)$ has finite increments in any point of the state space, namely that
\begin{equation}
\label{eq:finincr}
\sup_{y \in \Delta} \mathbbm E_y |H(Y(1)) - H(y)| < +\infty.
\end{equation}
Recall that $N(1)$ has density $\rho$ and has zero-mean and finite variance $\sigma^2$.

Consider $y = (u,v) \in \Delta$ such that $v\ge 1$. Then, since $0 < \lambda + \mu < \lambda$, we have that necessarily $y\in \Delta_1$. We obtain:
\begin{align*}
\mathbbm E_y |H(Y(1)) - H(y)| ={}& \int_{-\infty}^{\ell_1(v)} \left|\log \left( 1-\mu + \frac{\zeta + n} {\mu v} \right)\right| \rho(dn) \\
&{}+  \log v \int_{\ell_1(v)}^{\infty} \rho(dn),
\end{align*}
where $\ell_1(v) = \mu - \zeta - \mu(1-\mu)v$. If $N(1) > \ell_1(v)$, then by the definition of $H$ we have that $H(Y(1)) = 0$, in which case we are left with the second term (since $v\ge 1$ implies $\log v \ge 0$).

For a given $v$, the argument of the logarithm in the integral of the first term is less than $1$ for $n$ comprised between $\ell_0(v) := \mu^2v - \zeta$ and $\ell_1(v)$. Furthermore, it is lower bounded by $1-\mu + \frac {\ell_1(v)+\zeta}{\mu v} = \frac 1 v$, and, since $\log a \le a-1$ for $a>0$, we can write
\begin{align*}
\mathbbm E_y |H(Y(1)) - H(y)| & \le \int_{-\infty}^{\ell_0(v)} \log \left( 1-\mu + \frac{\zeta + n} {\mu v} \right) \rho (dn) \\
& \quad +  \log v \int_{\ell_0(v)}^{\ell_1(v)} \rho(dn) +  \log v \int_{\ell_1(v)}^\infty \rho(dn) \\
& \le -\mu + \frac \zeta {\mu v} + \frac 1 {\mu v}\int_{-\infty}^{\ell_0(v)} n \rho(dn) +  \log v \int_{\ell_0(v)}^\infty \rho(dn).
\end{align*}
All the terms in the sum above are bounded for $v>1$. The last term converges to $0$ as $v$ goes to infinity, since for large enough $v$ ensuring $\ell_0(v) > 0$ we have (by Chebyshev)
\begin{align*}
\log v \int_{\ell_0(v)}^\infty \rho(dn) &= \log v \cdot \PP{N(1)\ge \ell_0(v)} \\
& \le \log v \cdot \PP{|N(1)|\ge \ell_0(v)} \le \frac {\sigma^2 \log v} {\ell_0^2(v)} \to 0.
\end{align*}

Next consider $y = (u,v) \in \Delta$ such that $v < 1$. Thus $H(y) = 0$. We distinguish 4 cases: $M_1y \in D_1$, $M_1y \in D_2$, $M_1y \in D_3$, or $M_1y \in D_4$. Equivalently, the 4 cases write as
\begin{align*}
y\in \bar \Delta_1 &:= \{ (u, v) : u + v \ge 0, \ (\lambda+\mu)u + \lambda v > 0, v < 1 \}, \\
y\in \bar \Delta_2 &:= \{ (u, v) : u + v \ge 0, \ -r^* + \zeta < (\lambda+\mu)u + \lambda v \le 0 \}, \\
y\in \bar \Delta_3 &:= \{ (u, v) : u + v \ge 0, \ -r^* - \xi  < (\lambda+\mu)u + \lambda v \le -r^* + \zeta \}, \\
y\in \bar \Delta_4 &:= \{ (u, v) : u + v \ge 0, \ (\lambda+\mu)u + \lambda v \le -r^* - \xi \}.
\end{align*}

In the case $y \in \bar \Delta_1$, we have
\begin{align*}
\mathbbm E_y |H(Y(1)) - H(y)| ={} & \int_{-\infty}^{\ell_1(v)} \log [(1-\mu)v + \mu^{-1}(\zeta + n)] \rho(dn)\\
\le{} & (1-\mu)v\int_{-\infty}^{\ell_1(v)} \rho(dn) {+} \int_{-\infty}^{\ell_1(v)} ( \mu^{-1}(\zeta + n) - 1 ) \rho(dn).
\end{align*}
Both terms are bounded and converge to $0$ as $v$ goes to $-\infty$ (the first term again by Chebyshev, the second one by finiteness of the second moment).

In the case $y \in \bar \Delta_2$, it is straightforward to check that we necessarily have $v<0$. Define $\ell_2(y) := \mu - \zeta + \mu(\lambda+\mu)( u+v ) - \mu v$. We have that if $N(1) > \ell_2(y)$, then $H( Y(1) ) = 0$ by definition of $H(\cdot)$. Now since $u+v \ge 0$ and $\lambda+\mu>0$, it must be that
$$
- (\lambda+\mu)( u+v ) + v \le 0.
$$
Then,
\begin{align*}
\mathbbm E_y |H(Y(1)) - H(y)| & = \int_{-\infty}^{\ell_2(y)} \log [-(\lambda+\mu)(u+v) + v + \mu^{-1}(\zeta + n)] \rho(dn)\\
& \le \int_{-\infty}^{\ell_2(y)} [\mu^{-1}(\zeta + n) - 1] \rho(dn),
\end{align*}
which is clearly bounded.

The case $y \in \bar \Delta_4$ is closely related: Define $\ell_4(y) := \mu + \xi + \mu(\lambda+\mu)( u+v ) - \mu v$. We have that if $N(1) > \ell_4(y)$, then $H( Y(1) ) = 0$ by definition of $H(\cdot)$. Then, just like above,
\begin{align*}
\mathbbm E_y |H(Y(1)) - H(y)| \le \int_{-\infty}^{\ell_2(y)} [\mu^{-1}(-\xi + n) - 1] \rho(dn),
\end{align*}
which is again bounded.

In the case $y \in \bar \Delta_3$, it again holds that $v<0$. Define $\ell_3(y) := \mu - \zeta - (\lambda+\mu)(1-\mu)( u + v )$. We have that if $N(1) > \ell_3(y)$, then $H( Y(1) ) = 0$ by definition of $H(\cdot)$. Now since $u+v \ge 0$ and $\lambda+\mu>0$, it must be that
$$
\mu^{-1}(1-\mu)(\lambda+\mu)( u+v ) \le 0.
$$
Then,
\begin{align*}
\mathbbm E_y |H(Y(1)) - H(y)| & = \int_{-\infty}^{\ell_3(y)} \log \mu^{-1}[(1-\mu)(\lambda+\mu)(u+v) + r^* + n] \rho(dn)\\
& \le \int_{-\infty}^{\ell_2(y)} [\mu^{-1}(r^* + n) - 1] \rho(dn),
\end{align*}
which is again bounded.

Let us now examine the drift $\mathcal D H(y) = \mathbbm E_y(H(Y(1)) - H(y)$ in some point \mbox{$y\in\Delta_1\cap \{v\ge 1\}$}, which we denote by $\mathcal D H(v)$ by abuse of notation:
\begin{eqnarray*}
\mathcal D H(v) & = & \int_{-\infty}^{\ell_1(v)} \log \left( 1-\mu + \frac{\zeta + n} {\mu v} \right) \rho(dn) -  \log v \int_{\ell_1(v)}^{\infty} \rho(dn)\\
& = &  \log( 1-\mu ) \int_{-\infty}^{\ell_1(v)} \rho(dn) +  \int_{-\infty}^{\ell_1(v)} \log \left( 1 + \frac{\zeta + n} {\mu (1-\mu) v} \right) \rho(dn) \\
& & -  \log v \int_{\ell_1(v)}^{\infty} \rho(dn).
\end{eqnarray*}
Taking the limit as $v$ goes to infinity, we find that $\lim_{v\to \infty} \mathcal D H(v) = \log (1-\mu) > 0$. Indeed, each of the last two terms converge to $0$. In other words, for any $\epsilon > 0$, there exists a $v_0( \epsilon ) > 1$ which is such that for all $v > v_0$, $|\mathcal D H(v) - \log (1-\mu) | < \epsilon$. Taking a sufficiently small $\epsilon$, say $\epsilon_0 = \frac 1 2 \log (1-\mu)$, we obtain that
\begin{equation}
\label{eq:poslvl}
\forall v > v_0(\epsilon_0), \quad \mathcal D H(v) > 0.
\end{equation}

Define the set $C := \{ y = (u,v) : v \le v_0(\epsilon_0)\} \cap \Delta$. Then any $y$ in the complementary $\bar C := C^c \cap \Delta$ is such that $H(y)>\sup_{y_1\in C} H(y_1)$ and $\mathcal D H(y) > 0$. Using~(\ref{eq:finincr}) and~(\ref{eq:poslvl}) we can apply Theorem~11.5.2 from Meyn and Tweedie~\cite{meyn-tweedie} and conclude.
\endproof %\end{proof}

%%% Local Variables:
%%% mode: latex
%%% TeX-master: "grid"
%%% End:

\appendix
%\input{evapo}
%s5 ###
\section{Impact of delayed heating}
\label{sec-evapo}
%s5.1 ###
\subsection{With a constant coefficient of performance}
\label{sec-evapo-1}
Assume the appliance is a heating system used to maintain the temperature of a building at temperature $T(t)$. We use the model in Chapter III-E of MacKay et al.\cite{mackay2010sustainable} Let $\eps$ be the coefficient of performance of the appliance, $T(t)$ the room temperature, and $\theta(t)$ the outside temperature. Assume in this section that the coefficient of performance is constant. Assume the heating appliance consumes an amount of energy energy equal to $d(t)$ at time slot $t$; the obtained room temperature $T(t)$ is defined by the following equation
\be
  d(t) \eps = K( T(t)- \theta(t)) + C( T(t)-T(t-1)),
  \ee
where $K$ is the building's leakiness and $C$ its thermal inertia. Note that this equation is valid if we assume that the appliance is used for heating only and not for cooling. We make no particular assumption on $d(t)$, which typically depends on thermostat regulation by the building occupants.

Consider two scenarios. In the first scenario, the supply is always sufficient, and the appliance does not need to adapt (it is not throttled by demand-response). The energy consumed by the appliance is $D^{a}(t)$, the naturally occurring demand. Let $T^*(t)$ be the temperature obtained. Thus
\bear
D^{a}(t) \eps & = & K( T^*(t)- \theta(t)) + C( T^*(t)-T^*(t-1))\;\;\; t=1\dots\tau \label{eq-ev1}
\eear
In the second scenario, the energy supply is not sufficient at times $1\dots\tau-1$ and is sufficient at time $\tau$. The energy consumed at times $1\dots\tau-1$ when the appliance receives demand-response signals is $D^{a}(t){-}F(t)$ where $F(t)$ is the frustrated demand. The accumulated frustrated demand at time $\tau{-}1$ is
\be Z(\tau-1)=\sum_{t=1}^{\tau-1}F(t).\ee
The obtained temperature is $T(t)<T^*(t)$ for $t=1\dots\tau-1$. We assume that $T(0)=T^*(0)$.
Assume that at time $\tau$, we use the energy required to obtain the same temperature as in the first scenario; let  $D^a(\tau)+Z(\tau)$ be this amount of energy, i.e. $Z(\tau)$ is the amount of backlogged demand that remains at time $\tau$. We have thus
\begin{align}
\label{eq-ev-2a}
(D^{a}(t)-F(t)) \eps = &K( T(t)- \theta(t)) + C( T(t)-T(t-1)) \;\;\; t{=}1\dots\tau{-}1
\\
\label{eq-ev-2b}
(D^a(\tau)+Z(\tau))\eps = & K( T^*(\tau)- \theta(\tau)) + C( T^*(\tau)-T(\tau-1))
\end{align}
By summation of the last two equalities it comes:
\begin{align}
\!\lp\sum_{t=1}^{\tau} D^a(t) +Z(\tau)-Z(\tau-1)\!\rp \eps& = C\lp T^*(\tau)-T^*(0)\rp  \nonumber\\
&{}\quad  +K\!\lp \sum_{t=1}^{\tau-1} (T(t)-\theta(t)) + (T^*(\tau)-\theta(\tau))\!\rp
\label{eq-ev-11}
\end{align}
and by summation of \eref{eq-ev1},
\be
\lp\sum_{t=1}^{\tau} D^a(t)\rp \eps =
K\lp \sum_{t=1}^{\tau} (T^*(t)-\theta(t))\rp
+ C\lp T^*(\tau)-T^*(0)\rp.
\label{eq-ev-12}
\ee
Thus, by subtraction:
\bear
Z(\tau)-Z(\tau-1)
& = &
 - \frac{K}{\eps} \lp \sum_{t=1}^{\tau-1} (T^*(t)-T(t))\rp < 0.
 \label{eq-ev-fin1}
\eear
Therefore, for this simple model, the backlogged demand diminishes, i.e. the total energy consumption is reduced
if heating is delayed. In other words, there is positive evaporation.

%s5.2 ###
\subsection{With heat pumps}
\label{sec-evapo-2}
The model above assumes that the coefficient of performance $\eps$ is constant. For some systems, this does not hold as it might depend on the initial room temperature $T(t)$, the target room temperature $T^*(t)$ and the outside temperature $\theta(t)$. In particular, for heat pumps, the coefficient of performance is reduced when the differences $T^*(t)-T(t)$ and $T^*(t)-\theta(t)$ are large. To understand what happens with such systems, we revisit the second scenario above and, for simplicity, we consider the simpler case where all the demand is frustrated at times $1$ to $\tau-1$. Equations~(\ref{eq-ev-2a}) and (\ref{eq-ev-2b}) are now replaced by
  \begin{align*}
 0&= K( T(t)- \theta(t)) + C( T(t)-T(t-1)) \;\;\; t{=}1\dots\tau{-}1  \\
  (D^a(\tau)+Z(\tau))\eps'(\tau) &= K( T^*(\tau)- \theta(\tau)) + C( T^*(\tau)-T(\tau-1))
  \end{align*}
and $D^a(t) = F(t)$ for  $t=1,\dots,\tau-1$. Here we assume for simplicity that $\eps$ is maintained constant in scenario 1, but that the coefficient of performance is smaller for scenario 2, i.e. $\eps'(\tau)< \eps$. This is because the heat pump needs to produce water at time $\tau$ warmer than in scenario 1 to be able to reach the target temperature of the room, and if the outside temperature is low, the efficiency is less.  Summing these equations and combining with \eref{eq-ev-12} gives
\be
Z(\tau)-Z(\tau-1) = - \frac{K}{\eps} \lp \sum_{t=1}^{\tau-1} (T^*(t)-T(t))\rp + \lp 1- \frac{\eps'(\tau)}{\eps}\rp \lp D^a(\tau)+Z(\tau)\rp.
\ee
Compare with \eref{eq-ev-fin1}: the last term is positive, and if large, it might well be that $Z(\tau)>Z(\tau-1)$. i.e., there might be some negative evaporation.

%%% Local Variables:
%%% mode: latex
%%% TeX-master: "grid"
%%% End:

%\input{ode}
%s6 ###
\section{The ODE model}
\label{a-ode}
We can explicitly solve the associated ODE:
\begin{equation}
\dot x(t) = \sum_{i=1}^4\mathds{1}_{\{x(t) \in D_i\}} [\bar A_i x(t) + b_i], \ x(0) = x_0,
\end{equation}
where $\bar A_i := A_i - I_2$.
\eject

We obtain that the Cauchy problem has solution $x_i(t)$ in region $D_i$, where:
\begin{align*}
x_1(t) = & \left \{
\begin{array}{ll}
( I + \frac 1 \mu \bar A_1 ) ( x_1(t_1) + ( t - t_1 ) \zeta_0 ) &\\+
  \frac {e^{ -\mu (t-t_1)}} {\mu^2} \bar A_1 ( \zeta_0 - \mu x_1(t_1)
  ) - \frac 1 {\mu^2} \bar A_1 \zeta_0, & \mbox{when $\mu \ne 0$,}  \\
  \mbox{}&\\
  x_1(t_1) + (\bar A_1 x_1(t_1) + \zeta_0) ( t - t_1 ) + \frac 1 2
  \bar A_1 \zeta_0 (t-t_1)^2, & \mbox{when $\mu = 0$,}
\end{array}
\right.\\
x_2(t) = & \ ( I + \frac 1 {\lambda+\mu} \bar A_2 ) ( x_2(t_2)
+ ( t - t_2 ) \zeta_0 ) \\ & \ + \frac {e^{ -(\lambda+\mu) (t-t_2)}}
{(\lambda+\mu)^2} \bar A_2 ( \zeta_0 - (\lambda+\mu) x_2(t_2) )
-\frac 1 {(\lambda+\mu)^2} \bar A_2 \zeta_0, \\
x_3(t) = & \ \frac {e^{-(\lambda+\mu) (t-t_3) } } {(1-\lambda-\mu)(\lambda+\mu)} ( I + \bar A_3 ) ( (\lambda+\mu) x_3(t_3) - r^*_0 ) \\ & \ + \frac {e^{ -( t-t_3 )}} {1-\lambda-\mu} ( (\lambda+\mu) I + \bar A_3 )( r^*_0 - x_3(t_3) )
+ \frac {( \bar A_3 + ( 1+\lambda+\mu ) I )r^*_0} {\lambda+\mu}, \\
x_4(t) = & \ ( I + \frac 1 {\lambda+\mu} \bar A_4 ) ( x_4(t_4) - ( t - t_4 ) \xi_0 ) \\
& \ - \frac {e^{ -(\lambda+\mu) (t-t_4)}} {(\lambda+\mu)^2} \bar A_4 ( \xi_0 + (\lambda+\mu) x_4(t_4) ) + \frac 1 {(\lambda+\mu)^2} \bar A_4 \xi_0.
\end{align*}
We denoted by $t_i$ the instant when the system trajectory enters $D_i$.

\bibliographystyle{plain}

\end{document}